\documentclass[12pt]{article}

\usepackage{epsfig,amsfonts}

\def\bbox#1{{ \mbox{\boldmath $#1$}} }
\def\RP#1{\mathbf{R}\mathrm{P}^#1}

\def\Orth#1{\mathrm{O}(#1)}

\begin{document}

\title{Is trivial the antiferromagnetic $\RP{2}$ model\\
 in four dimensions?}

\author{H.~G.~Ballesteros$^{a}$,
J.~M.~Carmona$^{b}$, L.~A.~Fern\'andez$^{a}$, \\
V.~Mart\'{\i}n-Mayor$^{a}$, A.~Mu\~noz Sudupe$^{a}$ and
A.~Taranc\'on$^{b}$.}

\bigskip
\maketitle

\begin{center}
{\it a}~Departamento de F\'{\i}sica Te\'orica I, 
        Facultad de CC. F\'{\i}sicas,\\ 
{\it Universidad Complutense de Madrid, 28040 Madrid, SPAIN}\\
{\small e-mail: \tt hector, laf, victor, sudupe@lattice.fis.ucm.es}\\

{\it b}~Departamento de F\'{\i}sica Te\'orica,
Facultad de Ciencias,\\
{\it Universidad de Zaragoza, 50009 Zaragoza, SPAIN} \\
{\small e-mail: \tt carmona, tarancon@sol.unizar.es}\\
\end{center}

\begin{abstract}
We study the antiferromagnetic $\RP{2}$ model in four dimensions. We
find a second order transition with two order parameters, one 
ferromagnetic and the other antiferromagnetic. The antiferromagnetic
sector has mean-field critical exponents and a renormalized coupling
which goes to zero in the continuum limit. The exponents of the
ferromagnetic channel are not the mean-field ones, but the difference
can be interpreted as logarithmic corrections. We perform a 
detailed analysis of these corrections and conclude the triviality
of the continuum limit of this model. 
\end{abstract}

\vfill
DFTUZ/96/20 \hfill{hep-lat/9611003}

\newpage

\section{Introduction}
\label{INTRSection}

The non-perturbative formulation of non-asymptotically free,
interacting field theories in four dimensions is yet to be
accomplished. The conventional analysis, for $\lambda \phi^4$ and
$\Orth{N}$ theories, yields triviality in four
dimensions~\cite{TRIVIALIDAD1}.  That is, once the continuum limit is
taken, correlation functions factorice as the Wick's theorem
prescribes for the gaussian theory. A possible way, in order to obtain
a model with a non-trivial continuum limit, is to introduce 
antiferromagnetism (AFM). Gallavotti and Rivasseau have considered
AFM actions to change the ultraviolet limit of $\phi^4$ 
theories~\cite{GALLAVOTTI&RIVASSEAU}. From a Statistical Physics 
point of view,
a great variety of AFM models in three dimensions has been studied
to obtain different qualitative behaviour from that of the 
corresponding ferromagnetic (FM) 
models~\cite{REFS 6-7-8-9-10 O(4)}. In four dimensions, recent works have
studied the possibility of new universality classes if AFM is 
added~\cite{ISING AF-O(4)-POLONYI-SHAMIR}. 

The AFM $\RP{2}$ model has recently been studied in
three dimensions, due to its exotic properties~\cite{SHROCK,RP2D3}. For
instance, it has a disordered, unfrustrated ground
state. Even more, it seems to present a full breaking of
the action's $\Orth{3}$ symmetry~\cite{RP2D3}. Perturbative studies of
this Spontaneous Symmetry Breaking (SSB) pattern, yield the $\Orth{4}$
universality class~\cite{AZARIA}. 
If this prediction holds true in four dimensions, the fate
of the model is triviality. 

However, this theoretical prediction has been questioned
in three dimensions by Monte Carlo (MC) simulations~\cite{RP2D3}. Therefore,
the study of the triviality of this model in four dimensions is
very interesting. It also would help to enlighten the situation
in three dimensions. We will see, although, that a detailed analysis
of the MC simulation of this model indicates the triviality
of its continuum limit through the appearance of logarithmic corrections to
the divergences of the observables of the theory. A special form
of the Finite Size Scaling (FSS) analysis which includes logarithmic
corrections will be used to deal with these corrections.

We define the model and observables in section \S\ref{ModelSection},
where we also describe the techniques we have used to measure the
critical exponents.  The results of the MC simulation are presented in
section \S\ref{MCSection}. The model exhibits a phase transition at a
negative coupling with two independent order parameters.  One of this
channels, the staggered one, presents mean-field critical exponents,
but the other, ferromagnetic, presents deviations.  We show in
sections \S\ref{MCSection} and \S\ref{LCSection} how the discrepancies
with the mean-field behaviour can be interpreted as logarithmic
corrections.

\section{The model}
\label{ModelSection}

We shall consider the $\RP{2}\equiv $ S$^2$/Z$_2$ (real projective space) spin 
model in four dimensions. Our basic variable is a three component normalized
spin $\bbox{v}_i$, 
interacting through a gauge Z$_2$-invariant action. As a local 
symmetry cannot be broken, these are effectively $\RP{2}$ variables
(only the direction of the vectors is relevant).
We consider a hypercubic lattice, with first neighbour coupling:

\begin{equation}
S= \beta \sum_{<i j>} ({\mbox{\boldmath $v$}}_i\cdot \mbox{\boldmath
$v$}_j)^2 \quad , \quad 
\mathcal{Z}=\int\left(\prod_i{\rm d}\bbox{v}_i\right){\rm e}^{S}
\label{ACCION}
\end{equation}

The ferromagnetic (positive coupling) model presents a first order
transition at $\beta\approx~0.94$.
The ground state consists of spins parallel
or antiparallel to an arbitrary direction, and the SSB is
\hbox{SO(3)/SO(2)}. The analysis of the antiferromagnetic counterpart
is trickier, given the more complicated nature of the ground
state. Let us call a lattice site, labeled by $(x,y,z,t)$, even or odd
according to the parity of $x+y+z+t$. In the ground state,
every even/odd spin is parallel or antiparallel to an 
arbitrary direction, while odd/even spins lie randomly on the perpendicular
plane. The corresponding SSB is \hbox{SO(3)/SO(2)}, which calls for
the $\Orth{3}$ universality class. However, fluctuations induce an
interaction between spins on the randomly-plane sublattice. In three
dimensions, this seems enough to break the remaining $\Orth{2}$ 
symmetry~\cite{RP2D3}, and to change the Universality Class.

\subsection{Definition of observables}
\label{DEFSection}

The natural $\RP{2}$ variables are given by the traceless tensorial field
${\bf T}_i$

\begin{equation}
{\rm T}_i^{\alpha\beta}= v_i^\alpha v_i^\beta-
        \frac{1}{3}\delta^{\alpha\beta} ,
\label{TENSORFIELD}
\end{equation}

whose lattice Fourier transform will be represented by 
$\widehat{\bf T}$:
\begin{equation}
\widehat{\bf T}_{\mbox{\scriptsize \boldmath $p$}}=\sum_{\mbox{\scriptsize \boldmath $r$}}
\exp(-i{\mbox{\boldmath $p$}}\mathbf{\cdot}
{\mbox{\boldmath $r$}})\ {\bf T}_{\mbox{\scriptsize \boldmath $r$}}.
\end{equation}
We will work in a $L^4$ lattice with periodic boundary conditions.
We define two order parameters, according to the discussion of the 
ground state above, the
intensive staggered (ferromagnetic) magnetization, as the 
sums of tensors on even sites minus (plus) those on odd sites, or equivalently
\begin{equation}
{\bf M}_\mathrm{s}=\frac{1}{L^4}\widehat{\bf T}_{(\pi,\pi,\pi,\pi)} \quad,\quad ({\bf
M}=\frac{1}{L^4}\widehat{\bf T}_{(0,0,0,0)}).
\label{MAGS}
\end{equation}
As no spontaneous symmetry breaking can occur on a 
finite lattice, in a MC simulation one needs to 
measure $\Orth{3}$-invariant operators. For the magnetization and the
susceptibility, we define

\begin{equation}
M=\left\langle \sqrt{{\rm tr} {\bf M}^2}\right\rangle\quad,\quad
\chi=L^4 \left\langle {\rm tr}{\bf M}^2\right\rangle ,
\label{TUNMAGS}
\end{equation}
and analogously with the staggered observables. 

A very useful quantity for a triviality study is the
Binder cumulant. 
For this model, we define
\begin{equation}
V_M=\frac{5}{2}\left(\frac{7}{5} - 
\frac{\left\langle ({\rm tr} {\bf M}^2)^2\right\rangle}
{\left\langle {\rm tr} {\bf M}^2\right\rangle^2}\right),
\label{BINDER}
\end{equation}
which, in the infinite volume limit, becomes 1 in the broken phase
and 0 in the symmetric one.
The cumulant for the staggered magnetization is defined analogously.

Another very interesting quantity is the second momentum correlation  
length defined as ~\cite{LONGCORR}
\begin{equation}
\xi_L=\left(\frac{\chi/F-1}{4\sin^2(\pi/L)}\right)^{1/2},
\end{equation}
where $F$ is the mean value of the trace of $\widehat{\bf T}$ squared
at minimal momentum ($2 \pi/L$ in any of the four directions).  For $\xi_\mathrm{s}$ 
we use $\chi_\mathrm{s}$ and $F_\mathrm{s}$, analogously defined from $\widehat{\bf
T}$ at momentum $(2 \pi/L+\pi,\pi,\pi,\pi)$ and permutations.

The field theoretical definition of the renormalized coupling constant can now 
be introduced
\begin{equation}
g_\mathrm{R}=V_M \left(L/\xi_L\right)^d,
\label{RENCOUP}
\end{equation} 
where $d$ is the dimension of the lattice.
We will consider the renormalized couplings associated with
the two different sectors.

In addition, we measure the energy, which is needed for the 
spectral density method~\cite{FERRSWEND}, invaluable for extrapolating 
MC measures to a neighbourhood of the critical coupling.

\subsection{Standard Finite Size Scaling}
\label{FSSection}

To study critical exponents, we have used a method specially
suited to the measurements of anomalous dimensions~\cite{RP2D3},~\cite{O234}.
Let us consider the mean value of an operator $O$, measured in a
size $L$ lattice, at a coupling value $\beta$ in the critical
region. Let $t$ be the reduced temperature 
$(\beta-\beta_\mathrm{c})/\beta_\mathrm{c}$.
The standard FSS formula states that~\cite{BREZIN}
\begin{equation}
\langle O(L,t)\rangle=\langle O(t)\rangle
\mathcal{F}_O\left(s(L,t)\right)\, ,\quad 
s(L,t)\equiv\frac{L}{\xi(t)}\, , 
\label{FSSORIG}
\end{equation}
where $O(t)$ means $O(\infty,t)$ and $\mathcal{F}_O$ is a smooth function. 
We suppose that the values of
$L$ and $\xi(t)$ are large, so that we ignore scaling corrections 
in (\ref{FSSORIG}). Now, we have $\langle O(t)\rangle\sim t^{-x_O}$,
which is the definition of the critical exponent $x_O$, 
and $\xi(t)\sim t^{-\nu}$, and we
can write $L^{x_O/\nu}=\langle O(t)\rangle s^{x_O/\nu}$. This allows to
write (\ref{FSSORIG}) as
\begin{equation}
\langle O(L,t)\rangle=L^{x_O/\nu}\mathcal{G}_O(s).
\label{FSSDOS}
\end{equation}
Applying (\ref{FSSDOS}) to the correlation length, it  gives 
$\xi(L,t)=L\mathcal{G}_\xi(s)$, so that 
\begin{equation}
x(L,t)\equiv\frac{\xi(L,t)}{L}=\mathcal{G}_\xi(s),
\label{CAMBIOVAR}
\end{equation}
and $s=\mathcal{G}^{-1}_{\xi}(x)$. From eq. (\ref{FSSDOS}) we have
\begin{equation}
\langle O(L,t)\rangle=
L^{x_O/\nu}\mathcal{G}_O\left(\mathcal{G}_\xi^{-1}(x)\right)\equiv
L^{x_O/\nu}f_O(x),
\end{equation}
and we finish up with the useful expression
\begin{equation}
\left\langle O(L,t)\right\rangle=L^{x_O/\nu}f_O
\left(\frac{\xi(L,t)}{L}\right)+
\cdots,
\label{FSS}
\end{equation}
where the dots stand for possible scaling-corrections. Let us
denote 
\begin{equation}
Q_O=\frac{\langle O(rL,t)\rangle}
{\langle O(L,t)\rangle},
\label{COCIENTE}
\end{equation}
we can produce with it a sensible measure of critical exponents:
\begin{equation}
\left.Q_O\right|_{Q_\xi=r}=r^{x_O/\nu}+\cdots.
\label{CRITEXPS}
\end{equation}
Therefore, from simulations of lattice sizes $L$ and $rL$, we can
extract the critical exponent $x_O/\nu$ from the quotient (\ref{COCIENTE}),
measured at the point where one correlation length is $r$ times the other.

\section{The Monte Carlo simulation}
\label{MCSection}

To simulate the system,
we have used a standard three-hits Metropolis algorithm, with an 
uncorrelated change proposal, achieving approximately a 50\% of acceptance.
The lattice sizes have been
$L=4,6,8,10,12,16,20$ and 24. For the larger sizes, 20 and 24, we have combined
Metropolis with an overrelaxed update, described in Appendix 
\S\ref{OVRApp}, to decrease the autocorrelation time.
The overrelaxed algorithm is not able to decrease the dynamic
critical exponent $z$, but 
nevertheless we save total CPU time when compared with
the simple Metropolis simulation.
  
The runs have been distributed over several
Workstations. We display in Table~\ref{TAU} the 
integrated autocorrelation time for $\chi_{\mathrm s}$
and the number of measurements performed
for every lattice size. Every two measurements are separated by
10 sweeps, each consisting of either one Metropolis update or one
Metropolis plus three overrelaxed updates when we use the latter algorithm. 

\begin{table}
\begin{center}
\begin{tabular*}{12cm}{c@{\extracolsep{\fill}}cc} \hline
$L$	&	$\tau(\chi_{\mathrm s})$
	&	Measures $\left(\times 10^3~ 
\tau\left(\chi_{\mathrm s}\right)\right)$ \\ \hline \hline
4	&0.644(16)	&30	\\
6	&1.189(20)	&60	\\
8	&2.26(3)	&64	\\
10	&3.65(11)	&19	\\
12	&5.53(17)	&21	\\
16	&10.7(2)	&39	\\
20	&17.0(10)/2.31(11)	&5/5 \\
24	&25.5(10)/3.53(17)	&4/2 \\
\hline
\end{tabular*}
\vspace{3mm}
\end{center}
\caption{Total number of measures and the corresponding
integrated correlation times $\tau$ for $\chi_{\mathrm s}$. 
In the larger lattices
the data of the overrelaxed simulations (right) are separated from those
of Metropolis (left) by slashes.}
\vspace{8mm}
\label{TAU}
\end{table}

\subsection{About the order parameters}
\label{OPSection}

\begin{figure}[htb]
\centerline{\epsfig{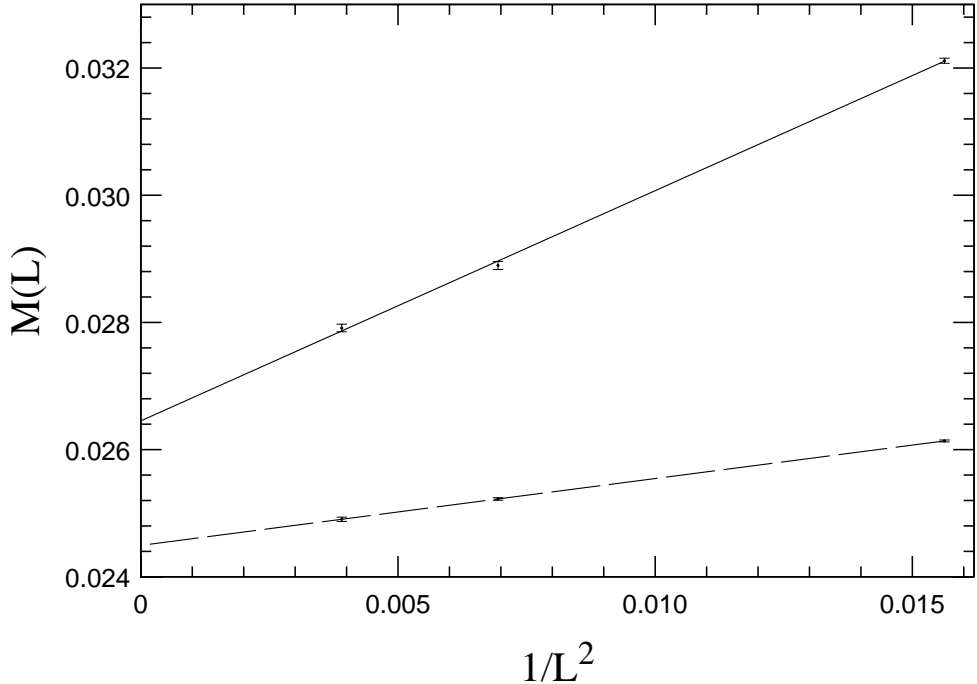}}
\caption{Asymptotic values of $M$ (straight line) and 
$M_{\mathrm s}/10$
(dashed line) from their values for $L=8,12$ and 16 at $\beta=-1.5$.}
\vspace{8mm}
\label{FIG:MAGN}
\end{figure}

The $\RP{2}$ model
presents a second order phase transition at $\beta\sim-1.34$.
The ferromagnetic and staggered magnetizations defined in equation
(\ref{TUNMAGS}) are zero below the transition.
To show that they are real order
parameters, we should ensure that they do not vanish in the broken
phase when $V\to\infty$. In Fig. \ref{FIG:MAGN} we plot the values of 
$M_{\mathrm s}$
and $M$ at $\beta=-1.5$ for the lattice sizes $L=8,12$ and 16. It is clear
that both magnetizations reach an asymptotic value different from zero
in the thermodynamical limit in the broken phase.

\subsection{Critical exponents}
\label{CREXPSection}

We have calculated the critical exponents $\nu$ and $\eta$
for the two different channels
using (\ref{CRITEXPS}), which yields
$x_\chi=\gamma$ for the susceptibility,
and $x_M=-\beta$ for the magnetization. To calculate $\nu$, we use  
$x_{\partial\xi/\partial\beta}=\nu+1$. All along this paper we shall
take $r=2$.

We obtain the anomalous dimension $\eta$ through the scaling relations:
\begin{equation}
(2-\eta)\,\nu=\gamma \quad , \quad 2\beta=\nu\,(d-2+\eta).  
\end{equation}
The resulting values for the $\eta$ exponent from these two relations, will be
denoted by $\eta_\chi$
and $\eta_M$ respectively.

As far as the exponent $\nu$ is concerned, we expect the same critical
exponent for both correlation lengths, the  ferromagnetic $\xi^{\mathrm{FM}}$
and the staggered $\xi$.  We have found that the measures for
$\xi$ are more accurate so we have used this variable as
correlation length.

\begin{figure}[htb]
\epsfig{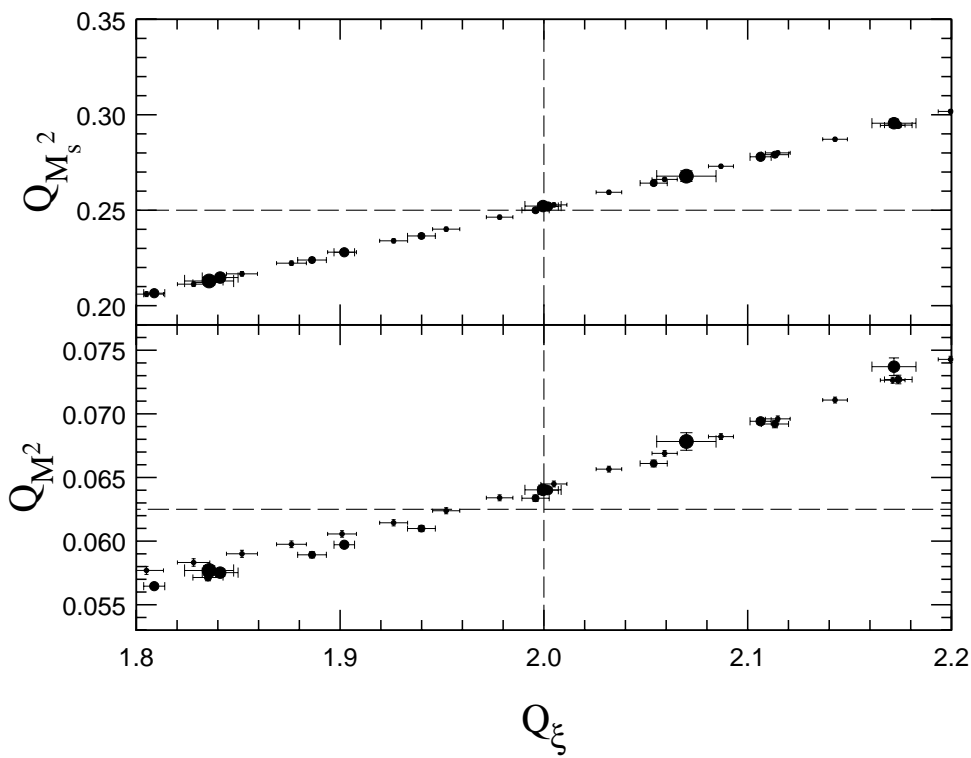}
\caption{Quotients of $M_{\mathrm {s}}^2$ and $M^2$ as a function
of the quotient of $\xi(L)$.  
The horizontal straight lines 
correspond to mean-field behaviour. The symbol sizes are proportional
to the lattice sizes.}
\vspace{8mm}
\label{FIG:M2}
\end{figure}

We plot in Fig. \ref{FIG:M2} an example of how this method works
in both channels. Notice that $Q_{M^2_{(\rm s)}}$ takes
the value $2^{\gamma_{(\rm s)}/\nu-d}$ when $Q_{\xi}=2$.

\begin{table}
\begin{tabular*}{\textwidth}
	{c@{\extracolsep{\fill}}ccccc} \cline{1-6}
\multicolumn{2}{c}{}	&	\multicolumn{2}{c}{Staggered}	
	& 	\multicolumn{2}{c}{Ferromagnetic} \\ \hline
$L_1,L_2$	& $\nu [0.5]$	  & $\eta_\chi [0]$ & $\eta_M [0]$ 
	& $\eta_\chi [2]$ & $\eta_M [2]$  \\ \hline \hline
4,8  & 0.527(8) & 0.009(4) & 0.008(4) & 1.959(8) & 1.963(9) \\
6,12 & 0.524(6) & 0.006(3) & 0.008(3) & 1.976(5) & 1.981(6) \\
8,16 & 0.512(4) & 0.008(2) & 0.005(2) & 1.968(3) & 1.973(3) \\
10,20& 0.491(7) & 0.013(4) & 0.001(4) & 1.964(6) & 1.969(6) \\
12,24& 0.496(9) & 0.007(5) & 0.009(5) & 1.960(8) & 1.965(7) \\
\hline
\end{tabular*}
\vspace{3mm}
\caption{Estimations for the critical exponents of the AFM $\RP{2}$ model.}
\vspace{8mm}
\label{TABLAEXPS}
\end{table}

The resulting exponents are shown in Table \ref{TABLAEXPS}.
 After the name of the exponents their mean-field values 
are shown in square brackets~\cite{RP2D3}.
The high accuracy reached on the measures of the $\eta$ exponents,
 is due in part to the
strong statistical correlation between $Q_\xi$ and $Q_{M^2}$.

We obtain a value for $\nu$ compatible with the mean-field prediction
as well as for the magnetic exponents of the staggered sector.
However, in the ferromagnetic channel, our exponents are close
but not compatible with those given by mean-field theory.

Another possible interpretation of these values is 
that they could be \textit{effective}
critical exponents due to the presence of strong logarithmic
corrections in the FM sector. 
To check this, let us suppose that there are logarithmic corrections only
in the susceptibility.
We do not take into account here the fact of possible logarithmic corrections
to $\xi$ as we use the values of the quotients measured at $r=2$ value.
This is an approximation that holds for large $L$.
We address to section \S\ref{LCSection} for a more complete treatment
of the logarithmic corrections. So that

\begin{equation}
\chi\sim L^m (\ln L)^{\bar m},
\end{equation}
\begin{equation} 
Q_\chi=\frac{\chi_{2L}}{\chi_L}=2^m\left(\frac{\ln 2L}{\ln L}
\right)^{\bar m}\equiv 2^m h_L^{\bar m},
\label{AJUSTEXI} 
\end{equation}
where $m=\gamma/\nu$. 
The effective exponents obtained with the 
standard FSS 
$m'_L=\ln Q_\chi/\ln 2$, can be written as
\begin{equation}
m'_L=m+\bar m\frac{\ln h_L}{\ln 2}.
\end{equation}
The fit discarding the 4,8 pair for the ferromagnetic susceptibility gives
\begin{equation}
m=0.09(9),\quad \bar m=-0.13(7),\quad 
\chi^2/\mathrm{d.o.f.}=0.04/2.
\label{FIT}
\end{equation}
The fit (\ref{FIT}) indicates that the exponents of the
ferromagnetic channel in Table \ref{TABLAEXPS} are compatible
with a mean-field situation with logarithmic corrections in the
susceptibility.

\subsection{Critical temperature} 
\label{CRTSection}

The thermodynamical critical temperature of our system can be
estimated from the crossing points of the Binder cumulants 
for the different lattice sizes~\cite{BINDER}.
To obtain $\beta_\mathrm{c}(\infty)$, we 
can  extrapolate according to the formula
\begin{equation}
\beta_\mathrm{c}(\infty)-\beta_\mathrm{c}(L)\approx L^{-1/\nu}.
\label{BETAFSS}
\end{equation}

It should be noted that (\ref{BETAFSS}) is only a first
approximation, because it does not include any logarithmic
corrections. In section \S\ref{LOGCORRSection} we will be able
to measure $\beta_\mathrm{c}(\infty)$ taking into account logarithmic 
corrections.  
However, it hardly modifies the value
of the critical coupling obtained with this method.

\begin{figure}[htb]
\epsfig{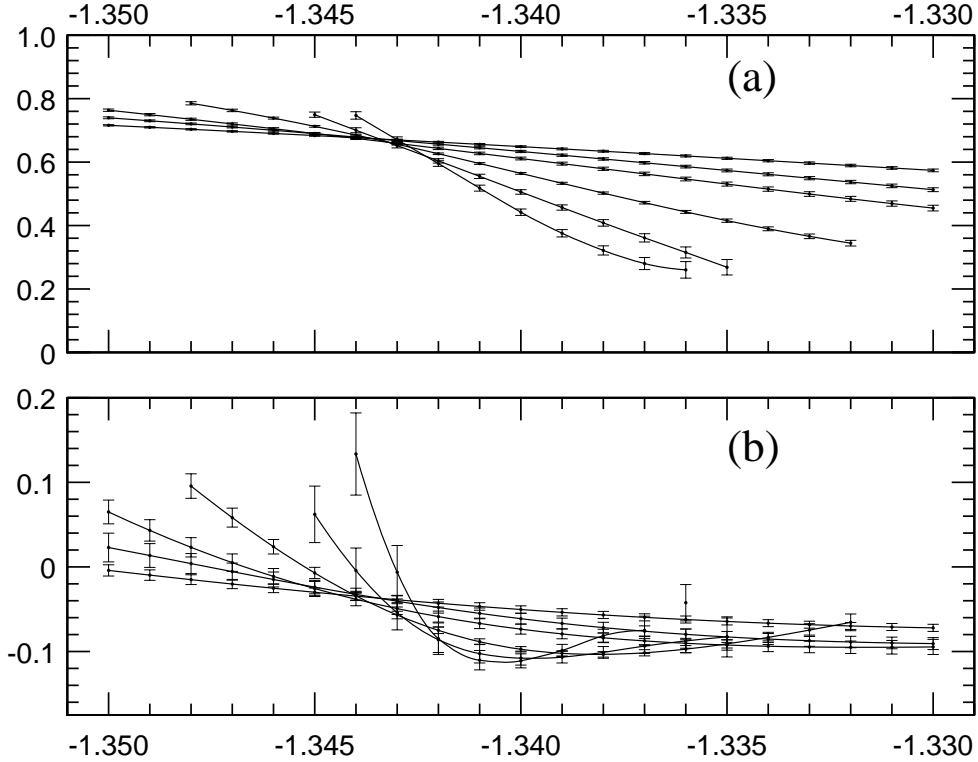}
\caption{Binder cumulants for both sectors, (a) staggered, (b) ferromagnetic.}
\label{FIG:CUMS}
\vspace{8mm}
\end{figure}

We show in Fig. \ref{FIG:CUMS} the Binder cumulants for the staggered
and ferromagnetic channels.

We have fitted the crossing points of the Binder cumulants 
of the $L=8$ lattice with 
lattices $L=12,16,20$ and 24. The results of the fit are
\begin{center}
\begin{tabular}{c@{\hspace{1cm}}c}
\bfseries Staggered & \bfseries Ferromagnetic \\
$\beta_\mathrm{c}(\infty)=-1.3426 (3)$ & $\beta_\mathrm{c}(\infty)=-1.3421 (6)$ \\ 
$\chi^2/{\rm d.o.f.}=1.2/2$ & $\chi^2/{\rm d.o.f.}=1.0/2$ \\
\end{tabular}
\end{center}

Both values are compatible, and, as we expect one transition
point, we take the value of $\beta_\mathrm{c}(\infty)$ with lower error,
that is
\begin{equation}
\beta_\mathrm{c}(\infty)=-1.3426(3).
\label{BETACRIT}
\end{equation}

Considering the crossing of the Binder cumulants of the $L=10$ lattice with 
the larger lattices scarcely change the numbers.
 
\section{Logarithmic corrections of the $\RP{2}$ model}
\label{LCSection}

As we have previously shown, the values obtained for the critical exponents for the four dimensional AFM $\RP{2}$ model,
are compatible with those predicted by mean-field plus  
logarithmic corrections.
They appear more clearly in the FM channel 
because in this sector there is no power-law divergence 
of the susceptibility, while in the staggered sector
the logarithmic divergence is added to a power-law one.
We need a modification of the standard FSS to include these corrections.

\subsection{FSS with logarithmic corrections}
\label{LOGSection}

Let us consider an observable $O(t)$ whose behaviour near the critical
point would have a logarithmic contribution
\begin{equation}
\langle O(t)\rangle\sim t^{-x}|\ln t|^{\bar x}.
\end{equation}
We will follow~\cite{KENNA} to take into account the logarithmic 
corrections: the scaling
variable $s(L,t)$ of (\ref{FSSORIG}) is now substituted by 
$\xi(L,0)/\xi(t)$:
\begin{equation}
\langle O(L,t)\rangle=
\langle O(t)\rangle\mathcal{F}_O\left(\frac{\xi(L,0)}{\xi(t)}\right).
\label{FSSNEW}
\end{equation}
Formula (\ref{FSSNEW}) coincides with (\ref{FSSORIG}) 
below the upper critical dimension, where $\xi(L,0)\sim L$, 
otherwise it can also take into 
account the logarithmic corrections to the finite volume 
correlation length. 

Let us suppose that, to leading order,
\begin{equation}
\xi(L,0)\sim L^{\hat{\alpha}} (\ln L)^{\hat{\beta}},
\label{XIL}
\end{equation}
where $\hat{\alpha}$ and $\hat{\beta}$ are two exponents that depend on the 
theory. For the O($N$) models, $\hat{\alpha}=1,~\hat{\beta}=\frac{1}{4}$
~\cite{BREZIN}.
The transition at finite $L$ takes place when $t$ is such that
$\xi(L,0)\sim\xi(t)$. If $\xi(t)\sim t^{-\nu}|\ln t|^{\bar\nu}$,
then, employing (\ref{XIL}),
\begin{equation}
t\sim L^{-\hat{\alpha}/\nu}(\ln L)^{(\bar\nu-\hat{\beta})/\nu}.
\label{BETALOGFSS}
\end{equation}

Below four dimensions $\hat{\alpha}=1$ and $\hat{\beta}=\bar\nu=0$. 
Now, we have
\begin{equation}
s(L,t)\equiv\frac{\xi(L,0)}{\xi(t)}\sim\frac{L^{\hat{\alpha}}
(\ln L)^{\hat{\beta}}}
{t^{-\nu}|\ln t|^{\bar\nu}},
\end{equation} 
so that making use of (\ref{BETALOGFSS}), and with a change of
variable similar to (\ref{CAMBIOVAR}), we obtain
\begin{equation}
\langle O(L,t)\rangle=
L^{\hat{\alpha} x/\nu}(\ln L)^{x/\nu(\hat{\beta}-\bar\nu)+\bar x}
f_O\left(\frac{\xi(L,t)}{L^{\hat \alpha}(\ln L)^{\hat \beta}}\right),
\label{FINALFSS}
\end{equation}
which is the equation analogue to (\ref{FSS}).
We follow now the same method as in the standard FSS case: we
compute the quotient
\begin{equation}
Q_O=\frac{\langle O(2L,t)\rangle}{\langle O(L,t)\rangle}=
2^{\hat{\alpha} x/\nu}\left(1+\frac{\ln 2}{\ln L}\right)^
{x/\nu(\hat{\beta}-\bar\nu)+\bar x}
\frac{f_O\left(\frac{\xi(2L,t)}{(2L)^{\hat \alpha}(\ln 2L)^{\hat
\beta}}\right)} {f_O\left(\frac{\xi(L,t)}{L^{\hat \alpha}(\ln L)^{\hat
\beta}}\right)}.
\label{COCBIS}
\end{equation}

Measuring $Q_O$ at the point $t_L$ where
\begin{equation}
\frac{\xi(2L,t_L)}{\xi(L,t_L)}=2^{\hat{\alpha}} h_L^{\hat{\beta}},
\label{MEDIDA}
\end{equation}

with $h_L\equiv 1+\frac{\ln 2}{\ln L}$, we find

\begin{equation}
Q_O(t_L)=2^{\hat{\alpha} x/\nu} h_L^{x/\nu(\hat{\beta}-\bar\nu)+\bar x}.
\end{equation}

This is the new expression that substitutes (\ref{CRITEXPS})
when there are logarithmic corrections, from it we can
extract the exponent $x/\nu$. We no longer have to measure
$Q_O$ where the quotient of correlation lengths is 2, but
instead where it equals $2^{\hat{\alpha}} h_L^{\hat{\beta}}$.

\subsection{Staggered channel: renormalized four-point \\
coupling}
\label{GRSection}

We proceed now to calculate the renormalized four-point coupling of our
theory. The limit we are interested in is
\begin{equation}
g_\mathrm{R}=\lim_{L\to\infty}g_\mathrm{R}(L,\beta_\mathrm{c}(\infty)).
\end{equation}

The evolution of $g_\mathrm{R}$ at the critical temperature with $L$ is
shown in Table \ref{TABLE:GR}, where we have used the value for
$\beta_\mathrm{c}(\infty)$ of (\ref{BETACRIT}). 

\begin{table}[htb]
\begin{center}
\begin{tabular*}{12cm}{c@{\extracolsep{\fill}}cc} \hline
$L$  	& Staggered  	& Ferromagnetic  \\ \hline \hline
8	& 3.16 (3)	& -0.19 (3)	\\
10	& 2.84 (4)	& -0.19 (5)	\\
12	& 2.61 (5)	& -0.21 (5)	\\
16	& 2.34 (4)	& -0.23 (5)	\\
20	& 2.08 (9)	& -0.23 (11)	\\
24	& 1.95 (13)	& -0.15 (17)	\\ \hline
\end{tabular*}
\end{center}
\vspace{3mm}
\caption{The renormalized four-point coupling 
$g_\mathrm{R}(L,\beta_\mathrm{c}(\infty))$.}
\vspace{8mm}
\label{TABLE:GR}
\end{table}

\begin{figure}[htb]
\centerline{\epsfig{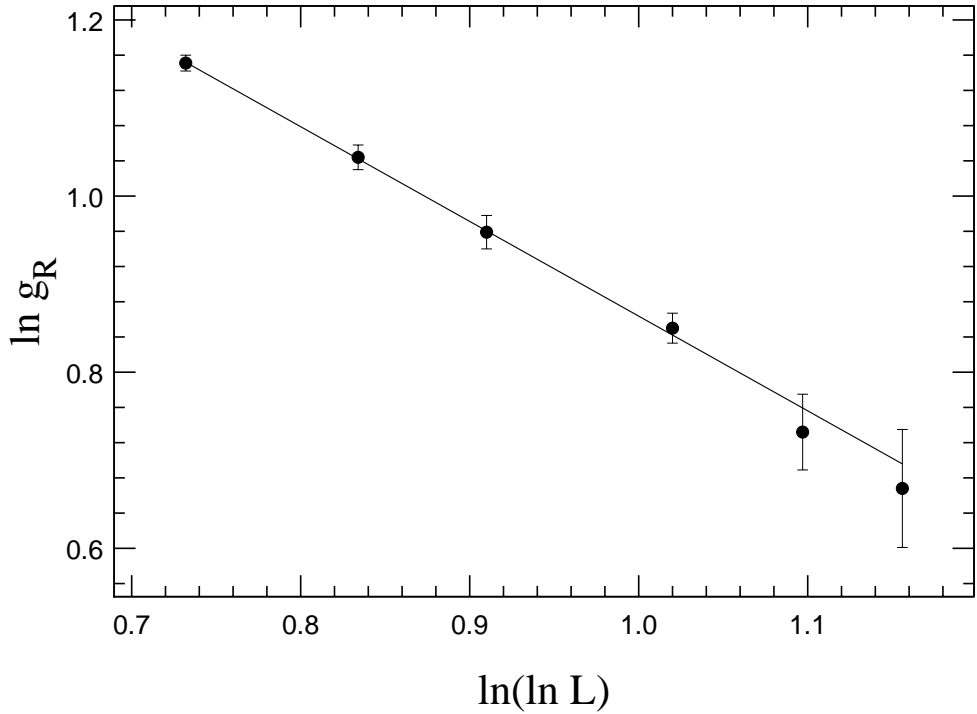}}
\caption{Fit of $g_\mathrm{R}(L)$ (staggered channel) 
at $\beta_{\mathrm {c}}(\infty)$.}
\label{FIG:FIT}
\vspace{8mm}
\end{figure}

When hyperscaling is violated by logarithms, 
$g_\mathrm{R}\sim|\ln t|^{-\bar\rho}$ \cite{FERNANDEZ}.
Let us apply the FSS formula (\ref{FSSNEW}):
\begin{equation}
g_\mathrm{R}(L,t)=g_\mathrm{R}(t)\mathcal{F}_{g_\mathrm{R}}\left(
\frac{L^{\hat{\alpha}}(\ln L)^{\hat{\beta}}}{t^{-1/2}|\ln t|^{\bar\nu}}
\right).
\label{GR(Lt)}
\end{equation} 
The scaling behaviour with $L$ at the critical point is
\begin{equation}
g_\mathrm{R}(L,0)=|\ln t|^{-\bar\rho}
\lim_{t\to 0}\mathcal{F}_{g_\mathrm{R}}\left(
\frac{L^{\hat{\alpha}}(\ln L)^{\hat{\beta}}}{t^{-1/2}|\ln t|^{\bar\nu}}
\right).
\label{GR(L)}
\end{equation}
We can eliminate the $t$ dependence in (\ref{GR(L)}) if
$\lim_{t\to 0}\mathcal{F}_{g_\mathrm{R}}(z)\sim z^{-\bar\rho/\bar\nu}$,
which makes the $|\ln t|^{-\bar\rho}$ factor disappear.
After that, we employ relation (\ref{BETALOGFSS}) and get finally
\begin{equation}
g_\mathrm{R}(L,0)\sim (\ln L)^{-\bar\rho},
\label{FIT2}
\end{equation}
which is also directly obtained from (\ref{FINALFSS}).
We can obtain $\bar\rho$ fitting the values of Table \ref{TABLE:GR} to the
functional form (\ref{FIT2}).
The fit for all lattice sizes, $L\geq 8$ yields (Fig. \ref{FIG:FIT})

\begin{equation}
\bar\rho=1.07(6), \quad \chi^2/\mathrm{d.o.f.}=0.8/4.
\label{FITRES}
\end{equation}

The result (\ref{FITRES}) implies triviality for the staggered
sector of the AFM $\RP{2}$ model in four dimensions. The renormalized
coupling goes to zero because of logarithmic corrections, exactly
in the same way as in the ferromagnetic $\Orth{N}$ models, 
for which $\bar\rho=1$
\cite{BREZIN&GUILLOU&ZINN-JUSTIN}.

The behaviour of the ferromagnetic channel is however rather
different from that of the staggered sector.
From the data of Table \ref{TABLE:GR} we cannot conclude
an asymptotic value for the renormalized coupling. 
We have seen above that the
logarithmic corrections are very strong in this channel, and
maybe for that reason, the renormalized coupling is very hard
to measure. To conclude about the triviality of this sector,
we will try to study in full detail the logarithmic corrections
of this channel, following the FSS analysis derived in 
section \S\ref{LOGSection}.

\subsection{Computation of the logarithmic corrections}
\label{LOGCORRSection}

\subsubsection{Correlation lengths}

To parameterize the logarithmic corrections, we
need a lot of, in principle, unknown exponents: 
$\hat{\alpha},\hat{\beta},\bar\nu,\ldots$ Therefore it will be necessary
to make a few assumptions about these exponents.

The most important exponents are $\hat{\alpha}$ and $\hat{\beta}$,
(\ref{XIL}), because we have to measure
at the points which satisfy (\ref{MEDIDA}). Assuming
a mean-field plus logarithmic corrections scenario,
we expect $\hat{\alpha}=1$. To find out the value of
exponent $\hat{\beta}$, we use the prediction (\ref{XIL}) for the 
correlation length in a finite lattice at $\beta_{\mathrm c}(\infty)$:

\begin{equation}
\ln\frac{\xi_L}{L}=C+\hat{\beta}\ln(\ln L).
\end{equation}

We have performed the fit for both correlation lengths, the
ferromagnetic ($\xi^{\mathrm{FM}}$), and the staggered ($\xi$) one. In
order to monitorize subleading effects, we have compared the fits with
$L\geq 8$, and $L\geq 10$. We have found that the fit for
$\xi^{\mathrm{FM}}$ is more stable with growing lattice sizes. In
Table~\ref{FIT_XI}, we show the fit parameters. The infinite volume
critical coupling, and the fit-parameter errors have been estimated
from the increment in one unit of the $\chi^2$ function. Comparing with
the previous determination of the critical coupling (\ref{BETACRIT}),
both determinations are consistent and of similar accuracy, although
logarithmic corrections to scaling were not considered previously.
Our value for the exponent $\hat \beta$ is consistent with the predicted
value for the O($N$) models, $\hat{\beta}=0.25$, specially the obtained
from the ferromagnetic correlation length.

\begin{table}[t]
\smallskip
\begin{center}
\begin{tabular}{|c|c|l|l|l|}\hline
  & Fit & \multicolumn{1}{c|}{$\chi^2/ \mathrm {d.o.f.}$} 
        & \multicolumn{1}{c|}{$\hat \beta$} 
        & \multicolumn{1}{c|}{$\beta_{\mathrm{c}}(\infty)$}      \\
\cline{1-5}
$\xi$ &$L_{\mathrm{min}}=8$  &2.1/3  &0.21(2)  & -1.3423(3)\\\cline{2-5}
  &$L_{\mathrm{min}}=10$    &0.3/2 &0.16(3)  & -1.3424(6)\\\hline  
$\xi^{\mathrm{FM}}$ &$L_{\mathrm{min}}=8$  &0.8/3 &0.22(4) & -1.3425(3)\\\cline{2-5}
  &$L_{\mathrm{min}}=10$    &0.2/2 &0.17(8)  & -1.3424(3)\\\hline  
\end{tabular}
\end{center}
\caption{Fits for the logarithmic corrections to the correlation
lengths at the critical point.}
\label{FIT_XI}
\medskip
\end{table}

\begin{figure}[htb]
\centerline{\epsfig{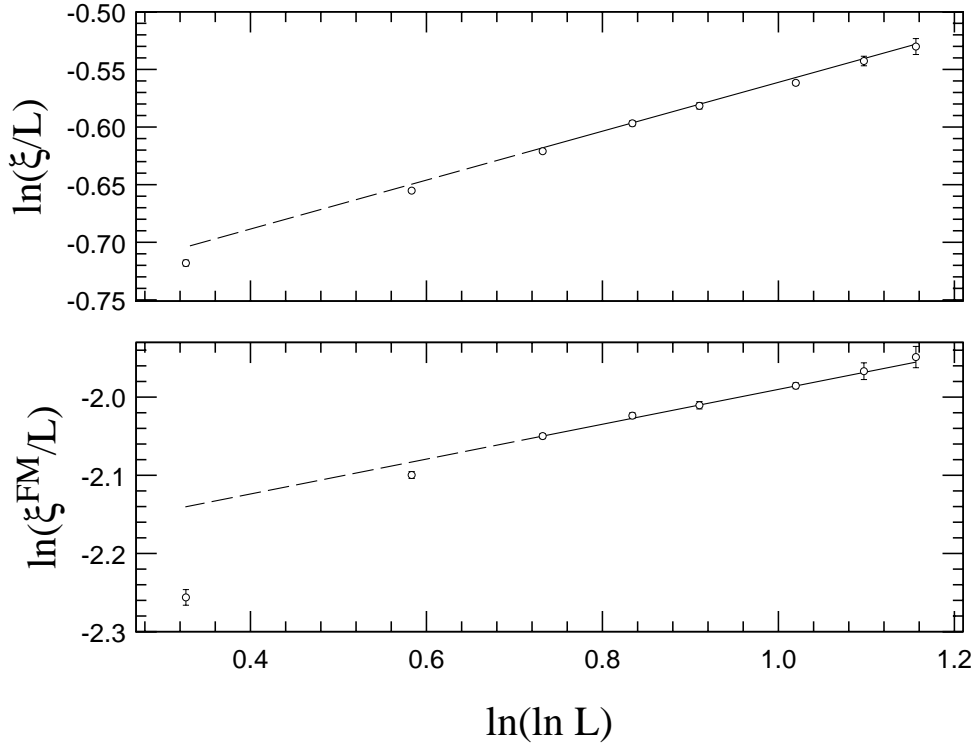}}
\caption{Determination of the exponent $\hat{\beta}$ of the FSS formulas
from the behaviour of $\xi^{\mathrm{FM}}$ and $\xi$ measured
at the mean values of $\beta_\mathrm{c}$ from the fits with $L\geq 8$.}
\vspace{8mm}
\label{FIG:BETA}
\end{figure}

\subsubsection{Magnetic operators}

We shall try to control other logarithmic 
corrections by making use of the results of section 
\S\ref{LOGSection}. 
From (\ref{COCBIS}) we know that
\begin{equation}
Q_\chi=2^{\hat{\alpha}\gamma/\nu}h_L^{\gamma/\nu(\hat{\beta}-\bar\nu)+\bar\gamma},
\label{COCCHI}
\end{equation}
when measuring the quotient at the point which 
verifies the condition (\ref{MEDIDA}).
The logarithmic corrections are given in terms of 
the unknown exponents $\bar\nu$ and $\bar\gamma$, but,
as we assume mean-field exponents, $\gamma=0$, $\nu=0.5$,
we can reduce (\ref{COCCHI}) to
\begin{equation}
\ln Q_\chi=\bar\gamma\ln h_L.
\end{equation}
In a similar way, if we take the magnetization,
\begin{equation}
Q_M=2^{-\hat{\alpha}\beta/\nu}
h_L^{-\beta/\nu(\hat{\beta}-\hat \nu)+\bar{\beta}},
\end{equation}
or (mean-field: $\beta=1$)
\begin{equation}
\ln Q_M=-2\ln 2+\bar\kappa\ln h_L,
\end{equation}
where $\bar\kappa=-2(\hat{\beta}-\bar\nu)+\bar{\beta}$.
Therefore, from every pair of lattices $L,2L$, we can 
obtain the exponents $\bar\gamma,\bar\kappa$. 
This is shown in Table \ref{EXPO}.

\begin{table}[htb]
\begin{center}
\begin{tabular*}{12cm}{c@{\extracolsep{\fill}}cc} \hline
$L_1,L_2$  & $\bar\gamma$  & $\bar\kappa$ \\ \hline \hline
4,8	& 0.45 (2)	& 0.23 (6)	\\
6,12	& 0.45 (1)	& 0.22 (6)	\\
8,16	& 0.49 (1)	& 0.24 (3)	\\
10,20	& 0.53 (2)	& 0.26 (1)	\\
12,24	& 0.52 (3)	& 0.26 (2)	\\ \hline
\end{tabular*}
\vspace{3mm}
\end{center}
\caption{Exponents of the logarithmic corrections of the FM
sector of the $\RP{2}$ model.}
\label{EXPO}
\medskip
\end{table}

Notice that the $\bar \gamma$ and $\bar \kappa$ values are very close
to their corresponding values for the magnetization in O(3) and O(4) models.
It should be understood that these values ($\bar \gamma \sim 0.5$ and 
$\bar \kappa \sim 0.25$) are calculated for the order parameter on
the fundamental (vectorial) representation of the O($N$) group.
However the critical exponents of the 
FM magnetization in the $\RP{2}$ model
and those of the order parameter 
in the tensorial representation for the O($N$) models 
are the same at the mean-field level. Let us remark that the 
$\bar \gamma \sim 2 \bar \kappa$ just means that $\langle M^2 \rangle \sim
{\langle M \rangle} ^2$.

This final result from the FM sector completes the
conclusion that we obtained after examining the renormalized coupling
of the staggered sector in \S\ref{GRSection}: the $\RP{2}$ model is
trivial due to the logarithmic corrections to the mean-field
behaviour. 

The question of the SSB pattern remains unsolved as the
ferromagnetic susceptibility is only logarithmically divergent. We recall
that the  power-law behaviour was crucial to check the symmetry 
breaking in three dimensions~\cite{RP2D3}.

\section{Conclusions}

We have examined the triviality question of the
four dimensional AFM $\RP{2}$ model, which presents a second 
order transition. A very interesting feature of this model is that it presents 
two different order parameters. A detailed study of these two sectors
reveals that the model has a trivial continuum limit.
We have been able to calculate explicitly the 
logarithmic corrections to the mean-field behaviour by means
of a FSS analysis also valid at the critical dimension of
the model.

\section{Acknowledgements}
We thank J~.L.~Alonso and J.~J.~Ruiz-Lorenzo 
for many enlightening discussions. This
work has been partially supported by CICyT AEN93-0604, AEN94-0218 and 
AEN95-1284-E.

\begin{appendix}
\section{The overrelaxed algorithm}
\label{OVRApp}

Overrelaxation is a local microcanonical update algorithm. 
It makes the maximum
change in the variable at a given point without modifying the energy.
We will now describe how this method works in our model. 
It is easy to see that

\begin{equation}
(\bbox{v}_i\cdot\bbox{v}_j)^2={\rm tr}{\bf T}_i{\bf T}_j+{1\over 3},
\end{equation}

so that the change ${\bf T}_i\to{\bf T}'_i$ must be such that  
the ${\rm tr}{\bf T}_i{\bf N}$ is conserved, where ${\bf N}$ is the
sum of the tensors at the neighbour points of $i$. If we set
${\bf T}'=R{\bf T}R^{-1}$, the conservation of the energy is then 
expressed by 
\begin{equation}
\left[R,{\bf N}\right]=0.
\label{COND1}
\end{equation}    
Besides, we must ensure that the new tensor belongs to $\RP{2}$, so
that the change in ${\bf T}$ is associated with a change in $\bbox{v}$:
$\bbox{v}'=R\bbox{v}$. As the vectors are normalized, this puts also the
condition of unitarity on the matrix $R$~\label{COND2}.

In order to fulfill these two conditions,
let us write ${\bf N}$ as ${\bf N}=U{\bf \Lambda}U^{-1}$,
where ${\bf \Lambda}$ is the matrix of eigenvalues of ${\bf N}$, and
define $C=U^{-1}RU$. Then, the updating conditions, written in terms of
the matrix $C$ are
\begin{equation}
\left[C,{\bf \Lambda}\right]=0 \quad , \quad C^{-1}=C^+.
\label{C-COND}
\end{equation}
As ${\bf \Lambda}$ is a diagonal matrix (\ref{C-COND}) implies that 
$C$ has to be
also diagonal, and 
$C^2={\bf 1}$, which means that its  three eigenvalues will be $\pm 1$.
We have reduced our updating process to a choice
of the matrix $C$. Here enters the second characteristic of the
overrelaxed algorithm: the change in the vector $\bbox{v}$ should be
maximum, which can be achieved by minimizing the value of the
squared scalar product 
\begin{equation}
\mathcal{A}=(\bbox{v}\cdot\bbox{v}')^2=(\bbox{v}\cdot R\bbox{v})^2=
(\tilde\bbox{v}\cdot C\tilde\bbox{v})^2,
\end{equation}
 where 
\begin{equation}
\tilde\bbox{v}=U^{-1}\bbox{v}\equiv(x_1,x_2,x_3).
\end{equation}

To do the update, we then have to take the three numbers $c_i=\pm 1$
that minimize the quantity
\begin{equation}
\mathcal{A}=(c_1x_1^2+c_2x_2^2+c_3x_3^2)^2.
\label{COMBINACIONES}
\end{equation}
To sum up, the overrelaxed algorithm consists of calculating the matrix
${\bf N}$ of nearest neighbours, its eigenvectors to obtain $U$,
and then looking for the minimum of the combinations in (\ref{COMBINACIONES}).

It is easy to see that this algorithm verifies detailed balance: if
we make $\bbox{v}\to\bbox{v}'\to\bbox{v}''$, so that
$\bbox{v}''=UC'\tilde\bbox{v}'$ and $\tilde\bbox{v}'=C\tilde\bbox{v}$,
we have to minimize
\begin{equation}
(\tilde\bbox{v}'\cdot C'\tilde\bbox{v}')^2=
(\tilde\bbox{v}\cdot CC'C\tilde\bbox{v})^2 ,
\end{equation}
and the last expression was minimized by $C$, so that $C=CC'C$ or $C'=C$.
Therefore $\bbox{v}''=\bbox{v}$.

\end{appendix}

\end{document}